\documentclass[twocolumn]{aastex61}

\usepackage{natbib,aas_macros} 
\usepackage{hyperref}
\usepackage{graphicx}	
\usepackage{amsmath,amssymb} 
\usepackage{xspace}
\usepackage{enumerate}
\usepackage{fp} 
\usepackage{lineno}
\usepackage{subfigure}

\newcommand{\ho}{$H_0$\xspace} 
\newcommand{\cchp}{\citetalias{cchp2proposal}\xspace}   
\newcommand{\hst}{\emph{HST}\xspace}
\newcommand{\ic}{IC\,1613\xspace}
\newcommand{\ngcfourfour}{NGC\,1448\xspace}
\newcommand{\ngconesix}{NGC\,1316\xspace}
\newcommand{\sne}{SN~Ia\xspace}

\newcommand{\mTRGBngcfourfour}{27.256}
\newcommand{\mTRGBngconesix}{27.400}

\newcommand{\mTRGBngcfourfourSTATerr}{0.021}
\newcommand{\mTRGBngconesixSTATerr}{0.0257}

\newcommand{\mTRGBngcfourfourSYSerr}{0.0002}
\newcommand{\mTRGBngconesixSYSerr}{0.008325}
 
\newcommand{\trgblum}{-4.00}
\newcommand{\trgblumstaterr}{0.03} 
\newcommand{\trgblumsyserr}{0.05}

\newcommand{\Iextinctionngcfourfour}{0.022} 
\newcommand{\Iextinctionngconesix}{0.032} 

\FPeval{\Iextinctionerr}{0.03/2}
\newcommand{\ZPerr}{0.02}
\newcommand{\EEerr}{0.02}
\newcommand{\Apcorrerr}{0.01}

\FPeval{\CalibrationErr}{ (\ZPerr^2+\EEerr^2+\Apcorrerr^2)^0.5 }

\FPeval{\ngcfourfourDM}{\mTRGBngcfourfour-\Iextinctionngcfourfour-\trgblum}
\FPeval{\ngconesixDM}{\mTRGBngconesix-\Iextinctionngconesix-\trgblum}

\FPeval{\ngcfourfourDist}{ 10^( (\ngcfourfourDM) /5)/100000 }
\FPeval{\ngconesixDist}{ 10^( (\ngconesixDM) /5)/100000 } 

\FPeval{\ngcfourfourmTRGBerrCMB}{ (\mTRGBngcfourfourSTATerr^2+\mTRGBngcfourfourSYSerr^2+\CalibrationErr^2)^0.5 }
\FPeval{\ngconesixmTRGBerrCMB}{ (\mTRGBngconesixSTATerr^2+\mTRGBngconesixSYSerr^2+\CalibrationErr^2)^0.5 }

\FPeval{\ngcfourfourCMBstaterr}{(\trgblumstaterr^2+\mTRGBngcfourfourSTATerr^2)^0.5  }
\FPeval{\ngcfourfourCMBsyserr}{(\trgblumsyserr^2+\mTRGBngcfourfourSYSerr^2+(\Iextinctionngcfourfour/2)^2+\CalibrationErr^2)^0.5  }
\FPeval{\ngcfourfourCMBerr}{ (\ngcfourfourCMBstaterr^2+\ngcfourfourCMBsyserr^2+\Iextinctionerr^2)^0.5 }
\FPeval{\ngconesixCMBstaterr}{(\trgblumstaterr^2+\mTRGBngconesixSTATerr^2)^0.5  }
\FPeval{\ngconesixCMBsyserr}{(\trgblumsyserr^2+\mTRGBngconesixSYSerr^2+(\Iextinctionngconesix/2)^2+\CalibrationErr^2)^0.5  }
\FPeval{\ngconesixCMBerr}{ (\ngconesixCMBstaterr^2+\ngconesixCMBsyserr^2+\Iextinctionerr^2)^0.5 }

\FPeval\ngcfourfourDistupperrdiststat{ 10^( (\ngcfourfourDM+\ngcfourfourCMBstaterr) /5)/100000 }
\FPeval\ngcfourfourDistlowerdiststat{ 10^( (\ngcfourfourDM-\ngcfourfourCMBstaterr) /5)/100000 }
\FPeval\ngcfourfourDiststaterr{ 0.5*(\ngcfourfourDistupperrdiststat - \ngcfourfourDistlowerdiststat) }
\FPeval\ngconesixDistupperrdiststat{ 10^( (\ngconesixDM+\ngconesixCMBstaterr) /5)/100000 }
\FPeval\ngconesixDistlowerdiststat{ 10^( (\ngconesixDM-\ngconesixCMBstaterr) /5)/100000 }
\FPeval\ngconesixDiststaterr{ 0.5*(\ngconesixDistupperrdiststat - \ngconesixDistlowerdiststat) }

\FPeval\ngcfourfourDistupperrdistsys{ 10^( (\ngcfourfourDM+\ngcfourfourCMBsyserr) /5)/100000 }
\FPeval\ngcfourfourDistlowerdistsys{ 10^( (\ngcfourfourDM-\ngcfourfourCMBsyserr) /5)/100000 }
\FPeval\ngcfourfourDistsyserr{ 0.5*(\ngcfourfourDistupperrdistsys - \ngcfourfourDistlowerdistsys) }
\FPeval\ngconesixDistupperrdistsys{ 10^( (\ngconesixDM+\ngconesixCMBsyserr) /5)/100000 }
\FPeval\ngconesixDistlowerdistsys{ 10^( (\ngconesixDM-\ngconesixCMBsyserr) /5)/100000 }
\FPeval\ngconesixDistsyserr{ 0.5*(\ngconesixDistupperrdistsys - \ngconesixDistlowerdistsys) }

\FPeval\ngcfourfourDisterr{(\ngcfourfourDiststaterr^2+\ngcfourfourDistsyserr^2)^0.5}
\FPeval\ngcfourfourDisterrPCent{(\ngcfourfourDiststaterr^2+\ngcfourfourDistsyserr^2)^0.5/\ngcfourfourDist*100}

\FPeval\ngconesixDisterr{(\ngconesixDiststaterr^2+\ngconesixDistsyserr^2)^0.5}
\FPeval\ngconesixDisterrPCent{(\ngconesixDiststaterr^2+\ngconesixDistsyserr^2)^0.5/\ngconesixDist*100}

\FPeval{\IextinctionngcfourfourROUNDED}{round(\Iextinctionngcfourfour,2)}
\FPeval{\IextinctionngconesixROUNDED}{round(\Iextinctionngconesix,2)}

\FPeval{\mTRGBngcfourfourROUNDED}{round(\mTRGBngcfourfour,2)}
\FPeval{\mTRGBngconesixROUNDED}{round(\mTRGBngconesix,2)}

\FPeval{\mTRGBngcfourfourSTATerrROUNDED}{round(\mTRGBngcfourfourSTATerr,2)}
\FPeval{\mTRGBngconesixSTATerrROUNDED}{round(\mTRGBngconesixSTATerr,2)}

\FPeval{\mTRGBngcfourfourSYSerrROUNDED}{round(\mTRGBngcfourfourSYSerr,2)}
\FPeval{\mTRGBngconesixSYSerrROUNDED}{round(\mTRGBngconesixSYSerr,2)}

\FPeval{\ngcfourfourmTRGBerrCMBROUNDED}{ round(\ngcfourfourmTRGBerrCMB,2) }
\FPeval{\ngconesixmTRGBerrCMBROUNDED}{ round(\ngconesixmTRGBerrCMB,2) }

\FPeval{\ngcfourfourDMROUNDED}{round(\ngcfourfourDM,2)}
\FPeval{\ngconesixDMROUNDED}{round(\ngconesixDM,2)}

\FPeval{\ngcfourfourCMBstaterrROUNDED}{ round(\ngcfourfourCMBstaterr,2)  }
\FPeval{\ngcfourfourCMBsyserrROUNDED}{ round(\ngcfourfourCMBsyserr,2) }
\FPeval{\ngcfourfourCMBerrROUNDED}{ round(\ngcfourfourCMBerr,2) }

\FPeval{\ngconesixCMBstaterrROUNDED}{ round(\ngconesixCMBstaterr,2)  }
\FPeval{\ngconesixCMBsyserrROUNDED}{ round(\ngconesixCMBsyserr,2) }
\FPeval{\ngconesixCMBerrROUNDED}{ round(\ngconesixCMBerr,2) }

\FPeval{\ngcfourfourDistROUNDED}{round(\ngcfourfourDist,1)}
\FPeval{\ngconesixDistROUNDED}{round(\ngconesixDist,1)}

\FPeval{\ngcfourfourDiststaterrROUNDED}{ round(\ngcfourfourDiststaterr,1) }
\FPeval{\ngconesixDiststaterrROUNDED}{ round(\ngconesixDiststaterr,1) }

\FPeval{\ngcfourfourDistsyserrROUNDED}{ round(\ngcfourfourDistsyserr,1) }
\FPeval{\ngconesixDistsyserrROUNDED}{ round(\ngconesixDistsyserr,1) }

\FPeval\ngcfourfourDisterrROUNDED{ round(\ngcfourfourDisterr,1) }
\FPeval\ngcfourfourDisterrPCentROUNDED{round(\ngcfourfourDisterrPCent,1)}

\FPeval\ngconesixDisterrROUNDED{ round(\ngconesixDisterr,1) }
\FPeval\ngconesixDisterrPCentROUNDED{round(\ngconesixDisterrPCent,1)}

\defcitealias{cchp2proposal}{CCHP}
\defcitealias{2016ApJ...832..210B}{Paper I}
\defcitealias{hatt17}{Paper II}
\defcitealias{2018ApJ...852...60J}{Paper III}
\defcitealias{hatt2018a}{Paper IV}

\shorttitle{TRGB distances to \ngcfourfour and \ngconesix}
\shortauthors{Hatt et al.}

\begin{document}

\title{\textit{The Carnegie-Chicago Hubble Program.} V. THE DISTANCES TO \ngcfourfour and \ngconesix\\ VIA THE TIP OF THE RED GIANT BRANCH\footnote{Based on observations made with the NASA/ESA \emph{Hubble Space Telescope}, obtained at the Space Telescope Science Institute, which is operated by the Association of Universities for Research in Astronomy, Inc., under NASA contract NAS 5-26555. These observations are associated with program \#13691.}}

\author[0000-0003-2767-2379]{Dylan~Hatt}\affil{Department of Astronomy \& Astrophysics, University of Chicago, 5640 South Ellis Avenue, Chicago, IL 60637}\email{dhatt@uchicago.edu}

\author{Wendy~L.~Freedman}\affil{Department of Astronomy \& Astrophysics, University of Chicago, 5640 South Ellis Avenue, Chicago, IL 60637}

\author{Barry~F.~Madore}\affil{Department of Astronomy \& Astrophysics, University of Chicago, 5640 South Ellis Avenue, Chicago, IL 60637}\affil{Observatories of the Carnegie Institution for Science 813 Santa Barbara St., Pasadena, CA~91101}

\author{In~Sung~Jang}\affil{Leibniz-Institut f\"{u}r Astrophysik Potsdam, D-14482 Potsdam, Germany}


\author[0000-0002-1691-8217]{Rachael L. Beaton}\altaffiliation{Hubble Fellow}\altaffiliation{Carnegie-Princeton Fellow}\affiliation{Department of Astrophysical Sciences, Princeton University, 4 Ivy Lane, Princeton, NJ~08544}

\author{Taylor~J.~Hoyt}\affil{Department of Astronomy \& Astrophysics, University of Chicago, 5640 South Ellis Avenue, Chicago, IL 60637}

\author{ Myung~Gyoon~Lee}\affil{Department of Physics \& Astronomy, Seoul National University, Gwanak-gu, Seoul 151-742, Korea}

\author{Andrew~J.~Monson}\affil{Department of Astronomy \& Astrophysics, Pennsylvania State University, 525 Davey Lab, University Park, PA 16802}

\author{Jeffrey~A.~Rich}\affil{Observatories of the Carnegie Institution for Science 813 Santa Barbara St., Pasadena, CA~91101}

\author{Victoria~Scowcroft}\affil{Department of Physics, University of Bath, Claverton Down, Bath, BA2 7AY, United Kingdom}

\author[0000-0002-1143-5515]{Mark~Seibert}\affil{Observatories of the Carnegie Institution for Science 813 Santa Barbara St., Pasadena, CA~91101}


\begin{abstract}

The \emph{Carnegie-Chicago Hubble Program} (CCHP) is re-calibrating the extragalactic \sne distance scale using exclusively Population II stars. This effort focuses on the Tip of the Red Giant Branch (TRGB) method, whose systematics are entirely independent of the Population~I Cepheid-based determinations that have long served as calibrators for the \sne distance scale. We present deep \emph{Hubble Space Telescope} imaging of the low surface-density and low line-of-sight reddening halos of two galaxies, \ngcfourfour and \ngconesix, each of which have been hosts to recent \sne events. Provisionally anchoring the TRGB zero-point to the geometric distance to the Large Magellanic Cloud derived from detached eclipsing binaries, we measure extinction-corrected distance moduli of $\ngcfourfourDMROUNDED\pm\ngcfourfourCMBstaterrROUNDED_{stat}\pm\ngcfourfourCMBsyserrROUNDED_{sys}$~mag for \ngcfourfour and $\ngconesixDMROUNDED\pm\ngconesixCMBstaterrROUNDED_{stat}\pm\ngconesixCMBsyserrROUNDED_{sys}$~mag for \ngconesix, respectively, giving metric distances of $\ngcfourfourDistROUNDED\pm\ngcfourfourDiststaterrROUNDED_{stat}\pm\ngcfourfourDistsyserrROUNDED_{sys}$~Mpc, and $\ngconesixDistROUNDED\pm\ngconesixDiststaterrROUNDED_{stat}\pm\ngconesixDistsyserrROUNDED_{sys}$~Mpc. We find agreement between our result and the available Cepheid distance for \ngcfourfour; for \ngconesix, where there are relatively few published distances based on direct measurements, we find that our result is consistent with the published \sne distances whose absolute scales are set from other locally-determined methods such as Cepheids. 
For \ngcfourfour and \ngconesix, our distances are some of the most precise (and systematically accurate) measurements with errors at 1.7 (2.8) \% and 1.6 (2.7) \% levels, respectively.

\end{abstract}


\section{Introduction} \label{sec:intro}

The tension in the value of \ho as determined by astrophysical methods \citep[for a review and recent updates see][]{2012ApJ...758...24F,2018arXiv180410655R} 
and indirect/modeling methods \citep[via the Cosmic Microwave Background; e.g.][]{2011ApJS..192...18K,2018arXiv180706209P} currently stands at 3.6-$\sigma$. Uncertainties and possible unknowns in the systematics of the Cepheid-based distance scale, such as the metallicity dependence of the period-luminosity zero-point  \citep[e.g.,][]{2004ApJ...608...42S,2013MNRAS.434.2418K,2013ApJ...777...79M} and the pervasiveness of source blending in extragalactic studies \citep[][]{2004ASPC..310...41M,2007A&A...473..847V}, have motivated the \emph{Carnegie-Chicago Hubble Program} (\cchp), which aims to re-calibrate the \sne extragalactic distance scale using an independent path based on Population (Pop) II stars \citep[a summary of the program is given in][Paper I]{2016ApJ...832..210B}. Since distances derived from Pop II stars are completely decoupled from the systematics encountered in the Pop~I path, they can provide insight into the current divide in the measurement of \ho. 

Pop~II stars are a natural---and in most cases, the preferred---substitute for Pop~I stars as distance indicators. Unlike Pop~I stars, which are typically found in the crowded and dusty disks of spiral and irregular galaxies, Pop~II stars can be universally found in the uncrowded, metal-poor, and virtually dust/gas-free outer halos of all galaxy types. The Pop II stars of focus for the \cchp are low-mass Red Giant Branch (RGB) stars at the end of the giant-branch evolutionary phase. These stars experience a rapid lifting of core-degeneracy, culminating in the He-flash and evolution into stable He-core burning at lower luminosities and bluer colors along the zero-age Horizontal Branch. This abrupt transition is sparked by a critical core mass that is weakly dependent on core metallicity and total star mass \citep{salaris_stellar_pop}.
Consequently, the tip of the RGB (TRGB) is a sharply-defined feature in Color-Magnitude Diagrams (CMDs), especially those of galaxy halos where few other stellar populations are present.  

This paper is a continuation in a series that presents TRGB distances to nine nearby galaxies containing a cumulative 12 \sne. Previously we have published a distance to the Local Group galaxy \ic \citep[][Paper II]{hatt17}, which, although it does not have a \sne on record, is an invaluable calibrator for the TRGB distance scale. We have furthered measured a distance to NGC\,1365 \citep[][Paper III]{2018ApJ...852...60J}, host to SN~2012fr, as well as distances to NGC\,4424, NGC\,4526, and NGC\,4536 \citep[][Paper IV]{hatt2018a}, which are host to SN~2012cg, SN~1994D, and SN~1981B, respectively.
The distances to \ic and NGC~1365 represent the extremes in distances for galaxies studied in the \cchp, approximately 730~kpc and 18.1~Mpc, which were measured to comparable precision owing to careful choice of field placement and the power of the TRGB method. 

In this work we present TRGB distances to a further two \sne host galaxies, \ngcfourfour and \ngconesix, using deep \emph{Hubble Space Telescope} (\hst) imaging of their halos. \ngcfourfour is an edge-on spiral galaxy that is host to the \sne event SN~2001el \citep{2001IAUC.7720....1M}, and \ngconesix is a lenticular galaxy in the Fornax Cluster that is host to the two \sne events SN~2006dd and SN 2006mr \citep{2006CBET..553....1M,2006CBET..729....1P}. These galaxies are believed to be comparable in distance to that of NGC\,1365, which makes them extremely valuable test cases for the most distant TRGBs in the \cchp sample. 

The paper is organized as follows: 
Section \ref{sec:data} describes the observations and photometry;  Section \ref{sec:trgb} presents the analysis of the TRGBs, including the estimation of measurement uncertainties and the determination of distances;  Section \ref{sec:dist_compare} places the distances measured here in context with previously-published estimates, including a Cepheid-based determination for \ngcfourfour; and Section \ref{sec:conc} provides a summary and the immediate impact of the results presented in this study.

\section{Data} \label{sec:data}

\begin{figure*}
\centering
\hspace{-100px}
\mbox{\includegraphics[angle=0,width=0.5\textwidth]{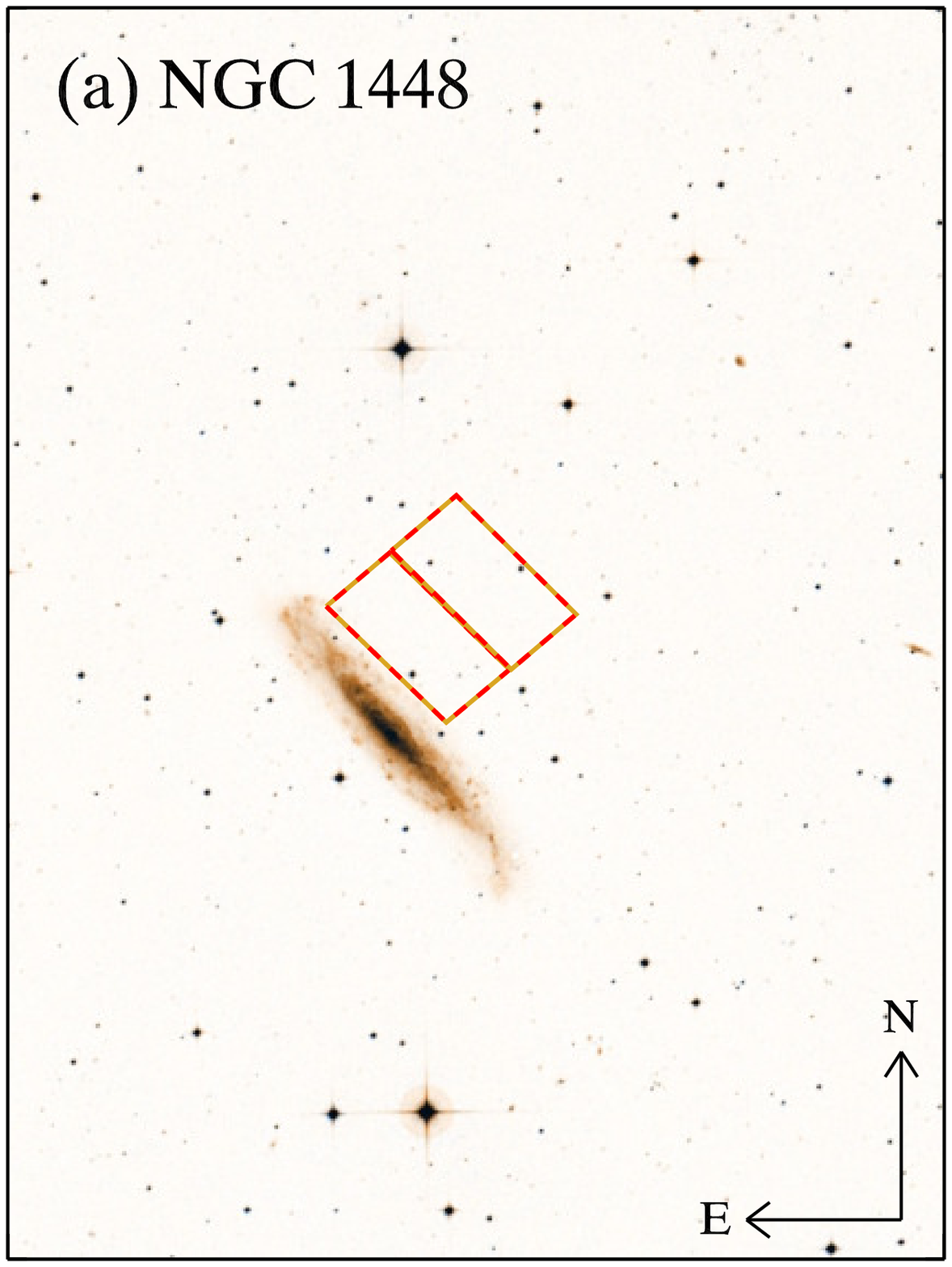}}
\hspace{-75px}
\mbox{\includegraphics[angle=0,width=0.5\textwidth]{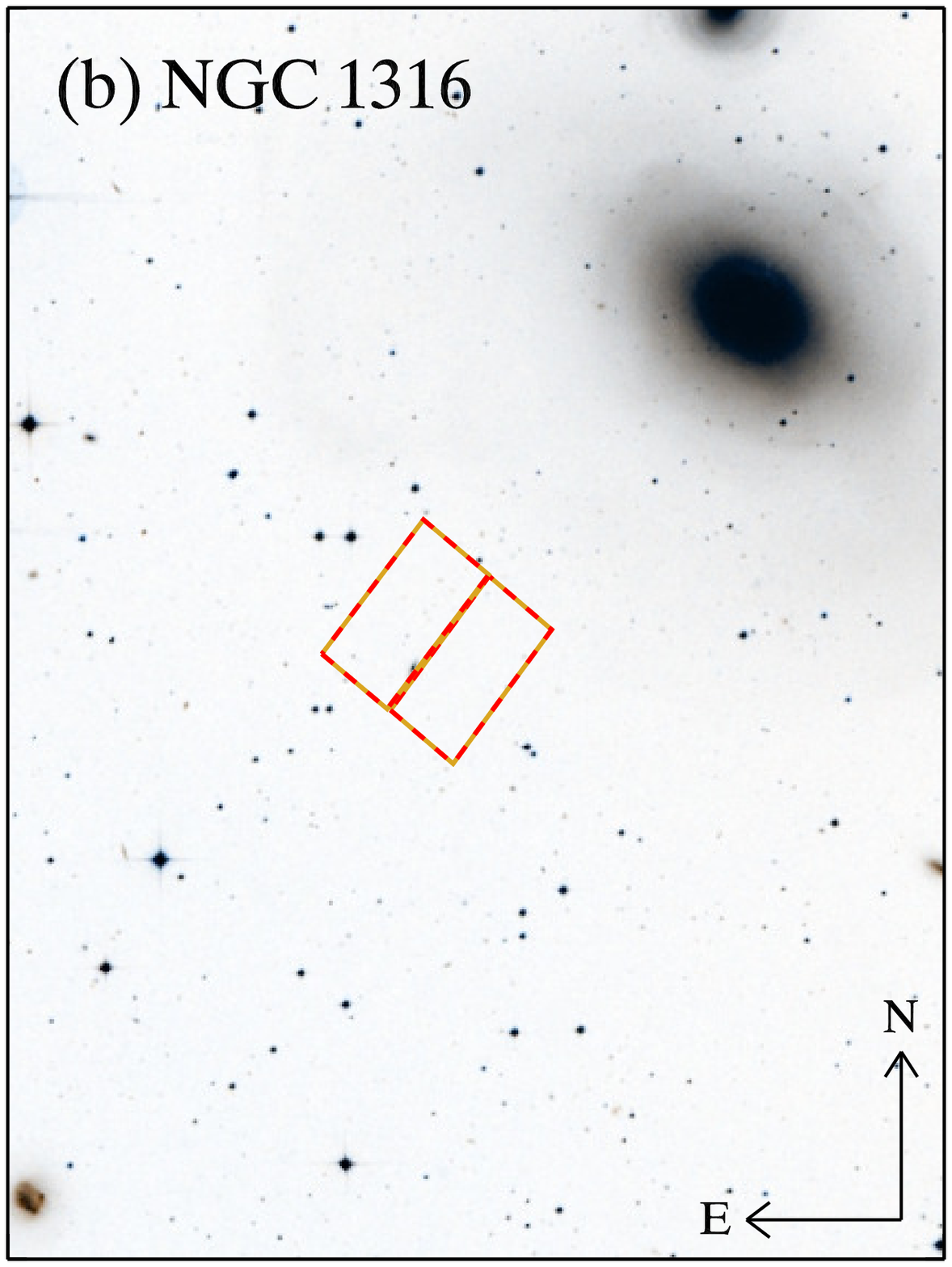}}
\hspace{-85px}
\vspace{-20pt}
\caption{Locations of $3\farcm37\times3\farcm37$ \hst ACS/WFC imaging (boxes) overlaid on an inverted color DSS image for (a) \ngcfourfour and (b) \ngconesix. 
\label{fig:imaging}}
\end{figure*}

\begin{deluxetable*}{cccccccc} 
\tabletypesize{\normalsize} 
\tablewidth{0pt} 
\tablecaption{ACS/WFC Observation Summary \label{tbl:obs_sum}} 
\tablehead{ 
\colhead{Target} &
\colhead{Dates} &
\colhead{Filters [No. obs~$\times$~exptime (s)]} &
\colhead{$\alpha$ (2000)} &
\colhead{$\delta$} (2000) &
\colhead{Field Size}
}
\startdata 
\ngcfourfour & 2015 Apr 03-12 & F606W $[7\times 1200]$, F814W [$15\times 1200$] & $03^h 44^m 27^s$ & $-44^\circ 36\arcmin 14\arcsec$  & $3\farcm37\times 3\farcm37\xspace$  \\
\ngconesix & 2015 Jul 03-11 & F606W [$12\times 1200$], F814W [$20\times 1200$] &  $03^h 23^m 13^s$ & $-37^\circ 19\arcmin 22\arcsec$ & $3\farcm37\times 3\farcm37\xspace$ \\
\enddata 
\tablecomments{See also \autoref{fig:imaging} for imaging coverage.} 
\end{deluxetable*} 

\subsection{Observations}

We have made use of the \hst Advanced Camera for Surveys using the Wide-Field Channel (ACS/WFC) \citep[Program \#13691,][]{cchp2proposal}, whose specifications and field selection criteria are described in \citetalias{2016ApJ...832..210B}. In short, these include placement along galaxies' minor axes to maximize the number of halo stars while avoiding disks and tidal structures. \autoref{fig:imaging} displays the imaging coverage for this study, and \autoref{tbl:obs_sum} provides a summary of the observations. 

Observations were taken mid-2015 over a span of 8-9 days for each galaxy. For \ngcfourfour, 4 and 8 orbits were devoted to imaging in the F606W and F814W filters, respectively; for \ngconesix, 6 and 10 orbits were devoted to the respective filters. Exposure durations of $\sim1200$~sec for each filter were set in order to achieve a signal-to-noise of 10 at the anticipated magnitude of the TRGB. The images used in the analysis are the FLC type, which are calibrated, flat-fielded, and CTE-corrected. We further corrected the images by multiplying by their Pixel Area Maps\footnote{\url{http://www.stsci.edu/hst/acs/analysis/PAMS}} to account for geometric distortions in the ACS/WFC camera.

\subsection{Photometry}

\begin{deluxetable*}{ccccccccc} 
\tabletypesize{\normalsize} 
\tablewidth{0pt} 
\tablecaption{Average measured aperture corrections at 0\farcs5\label{tbl:phot_cal}} 
\tablehead{ 
\colhead{Target} &
\multicolumn{2}{c}{CCD1}  &
\multicolumn{2}{c}{CCD2}   \\
\colhead{} &
\colhead{F606W (No.)} & \colhead{F814W (No.)} &
\colhead{F606W (No.)} & \colhead{F814W (No.)}
}
\startdata 
\ngcfourfour & $-0.16(14)$ & $-0.14(33)$ & $-0.13(8)$ & $-0.12 (33)$\\
\ngconesix & $-0.13 (15)$ & $-0.09 (24)$ & $-0.13 (26)$ & $-0.08 (50)$\\
\enddata 
\tablecomments{Parentheses are the total number of bright, isolated stars used in computing the average aperture corrections. Approximately equal numbers of stars are available in each image for aperture corrections on a frame-by-frame basis.} 
\end{deluxetable*}

Photometry was performed identically to that described in Section 2 of \citetalias{hatt17} and \citetalias{2018ApJ...852...60J}. In brief, photometry was carried out using the \textsc{DAOPHOT} suite of software \citep{1987PASP...99..191S} using Point-Spread-Functions (PSFs) that were obtained through the \hst Tiny Tim PSF modeling software \citep{2011SPIE.8127E..0JK}. Instrumental photometry was filtered for point sources through simple magnitude error and \textsc{DAOPHOT} `chi' and `sharpness' cuts: $\mathrm{F814W}_\mathrm{err} < 0.015+0.003\cdot \exp\left(\mathrm{F814W}-23.75\right)$,
$\mathrm{chi < 1.8}$, and $-0.3 < \mathrm{sharp} < 0.2$.

The calibration of instrumental magnitudes to the Vegamag system then followed the prescription in \citet{2005PASP..117.1049S}. We refer the reader to the detailed assessment of possible systematic uncertainties in the calibration process in Section 2 of \citetalias{2018ApJ...852...60J}, which include considerations of the uncertainty in Encircled-Energies for a given aperture size and in the flux of Vega itself. The photometric zero-points used in the following analysis, subject to change at future dates via the online calculator\footnote{\url{https://acszeropoints.stsci.edu/}}, are 26.405 mag for F606W and 25.517 mag for F814W, accessed on 2018-06-11. While calibration constants like the photometric zero-points are largely unchanged between the works in the \cchp because they belong to a single \emph{HST} observing cycle, unique here are the measured aperture corrections at $0\farcs5$ whose averages we list in \autoref{tbl:phot_cal}. In practice, these aperture corrections vary between exposures for a given filter at the 0.03-0.05~mag level. As per \citet{2005PASP..117.1049S}, they are applied in the following analysis on a frame-by-frame basis, but we have found in the previous papers in this series, as well as in the current work, that the TRGB science result is unchanged to within the reported uncertainties if using only the average.

\subsection{Color-magnitude diagrams}\label{sec:cmds}

\begin{figure*}
\centering
\hspace{-15px}
\mbox{\includegraphics[angle=0,width=0.45\textwidth]{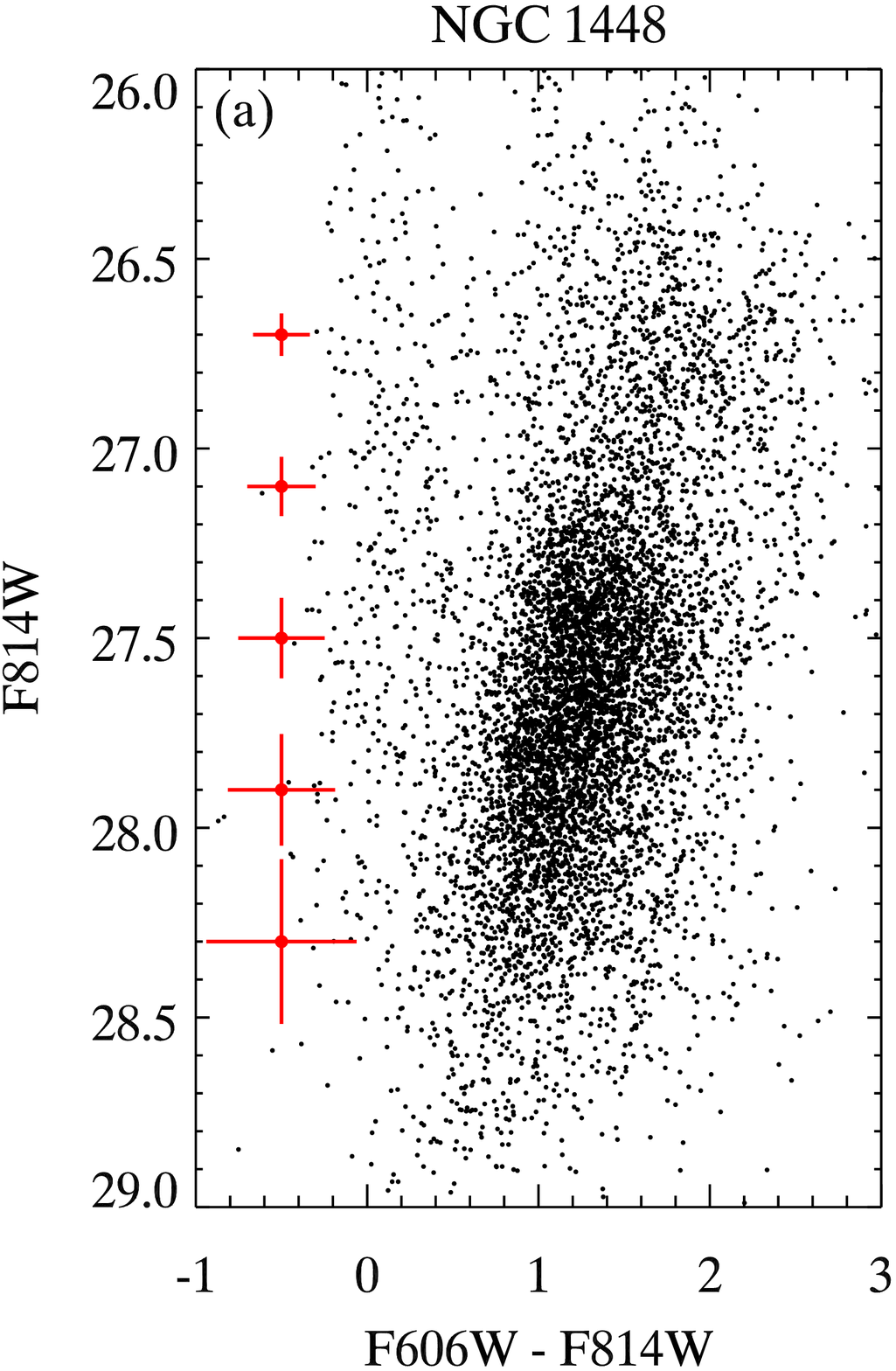}}
\hspace{-15px}
\mbox{\includegraphics[angle=0,width=0.45\textwidth]{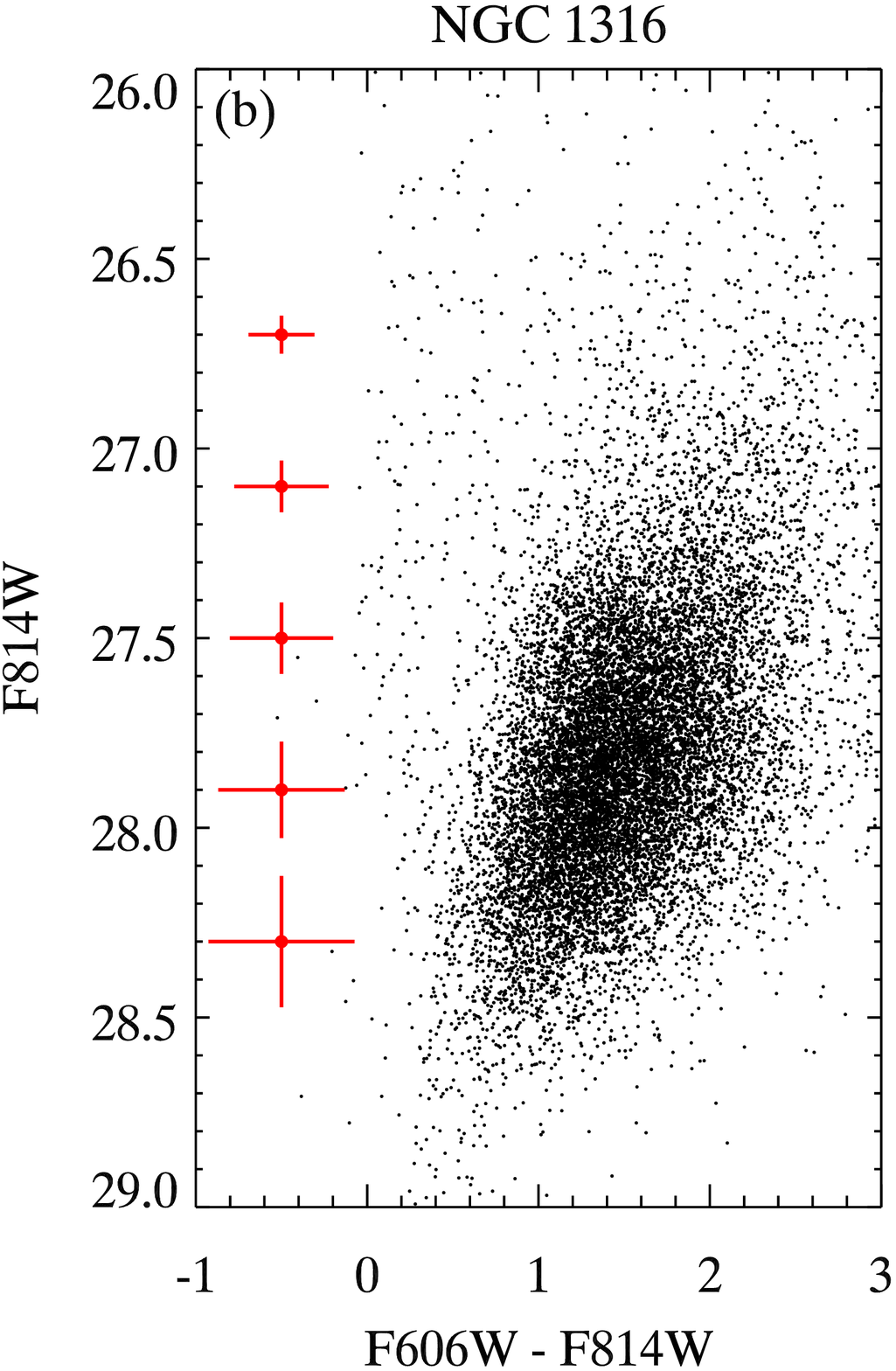}}
\hspace{-15px}
\vspace{-10pt}
\caption{Color-magnitude diagrams of the galaxies \ngcfourfour and \ngconesix. 
Likely TP-AGB stars are visible above the Red Giant Branch, although an abrupt jump in the luminosity function of each CMD, shown later in Section \ref{ssec:dist}, still clearly marks the location of the TRGB. Median magnitude and color error bars are plotted alongside the CMDs.
\label{fig:cmds}}
\end{figure*}

The calibrated photometry is presented in the form of CMDs in \autoref{fig:cmds}. Error bars show sample median color and magnitude uncertainties. Each CMD shows a dominant halo component, i.e. an RGB, mixed with likely Thermally-Pulsating Asymptotic Giant Branch (TP-AGB) stars as well as early-type AGB (E-AGB) stars. Hereafter, we often refer to the two AGB classes as a single `AGB' component that spans the full magnitude range of the RGB and extends brighter than the tip of the RGB.
In Sections \ref{ssec:artstar_lum}--\ref{sec:completeness_crowding}, we address the possibility of contamination in our measurement of the TRGB by AGB sources as well as assess the level of photometric completeness and crowding/blending through extensive artificial star tests.
In addition to an RGB/AGB component, each CMD contains a muted blue component centered near $\mathrm{F606W}-\mathrm{F814W}=0.0$. These objects are possibly young and massive stars, but they also occupy a color-magnitude space populated by components of background galaxies that pass photometry selection criteria. In either case, we explore the use of a color-magnitude selection cut to further isolate the RGB/TRGB in Section \ref{ssec:dist}.

Each CMD contains a large enough sample of stars such that the RGB is well-populated. \cite{2009ApJ...690..389M} showed that $\sim400$ stars were needed in the first magnitude below the TRGB to achieve $\pm0.1$~mag precision in the measurement of the TRGB. For these galaxies, the counts are over an order of magnitude larger, suggesting that the statistical (random) uncertainty in the TRGB measurement will be small. We note, however, that compared to higher signal-to-noise CMDs like that of IC\,1613 in \citetalias{hatt17}, the TRGB for each galaxy may be not as visually sharp due to larger photometric uncertainties. Then, in the case of distant galaxies such as those presented here, a precise measure of the TRGB magnitude is made possible through large population counts rather than a small number of well-measured, individual stars.

\section{The Tip of the Red Giant Branch} \label{sec:trgb}

\subsection{Background}

The TRGB is marked by a jump in a galaxy's luminosity function between the foreground/AGB populations and the RGB. In the $I$-band (or more specifically its \hst equivalent, F814W),
the observed TRGB is remarkably fixed in brightness for observations of metal-poor stars, such as the galaxy halo imaging presented here. Note also that color-magnitude corrections for higher-metallicity stars have recently been calibrated empirically for ACS/WFC filters by \citep{2017ApJ...835...28J}. In the case of such halo observations, where only AGB and RGB components are expected, it follows that the location of the TRGB is defined by the point of greatest change in the one-dimensional transition in star count (as a function of magnitude) between the two populations.

The simplicity of the TRGB method in the $I$-band allows for the application of basic yet robust tools to measure this point of transition. An overview of the \cchp approach to the TRGB method, as well as comparisons to existing methods, are given in \citetalias{hatt17} and \citetalias{2018ApJ...852...60J}. In brief, an edge-detector measures the first-derivative of finely binned and smoothed foreground/AGB and AGB/RGB luminosity functions, $[-1,0,+1]$, and returns a maximum response at the location of the TRGB. A common alternate method to measuring the TRGB is the simultaneous fit of the slopes of the AGB and RGB luminosity functions via a Maximum Likelihood estimator \citep[often cited is][]{2006AJ....132.2729M}. In the appendices of \citet{hatt17} and \citet{2018ApJ...852...60J}, we compare the performance of the most common edge-detection methods for galaxies IC\,1613 and NGC\,1365 and find no statistically significant difference in their measured TRGB values. Since IC\,1613 and NGC\,1365 represent some of the nearest and furthest galaxies in the \cchp sample, the equal performance of the edge-detection methods suggests that the edge-detection methodology applied here is suitable for all other \cchp galaxy targets as well.

In the following sub-sections, we describe how the parameters for the smoothing of the data are chosen in order to optimize the measurement of the TRGB for each galaxy.

\subsection{Artificial stars and luminosity functions}\label{ssec:artstar_lum}

We generate artificial AGB and RGB star populations, consistent with the previous papers in this series, in order to model the observed luminosity function around the TRGB as closely as possible. We adopt AGB and RGB population slopes of 0.1~dex and 0.3~dex, respectively. The RGB luminosity function begins at the input TRGB value and extends 1 magnitude fainter. The input TRGBs have been adjusted for the galaxies at hand, which we have set to the approximate and preliminary TRGB measurements of 27.3~mag and 27.5~mag for \ngcfourfour and \ngconesix, respectively. The AGB luminosity function begins 1 mag brighter than the input TRGB and extends the entire range of the RGB luminosity function. The relative counts of RGB:AGB stars is set to 4:1 at the TRGB to be consistent with recently observed population statistics in local galaxies \citep[see e.g.][among others]{rose_2014}. Stars were randomly assigned colors from a uniform distribution between $1.0 \leq \mathrm{F606W}-\mathrm{F814W} \leq 1.5$ to model the approximate widths of the observed CMDs. For a given iteration, $2000$ stars are sampled from these distributions and placed into each CCD. Photometry is performed identically to before, and the measured magnitudes are recorded. The process of generating and measuring artificial stars is repeated 250 times for each CCD, resulting in a cumulative $500,000$ simulated stars for each galaxy. The input and measured (output) luminosity functions for each galaxy are shown in \autoref{fig:artstars}a. 

\subsection{Photometric completeness and crowding}\label{sec:completeness_crowding}

The large number of artificial stars allows us to assess the quality of our photometry, including completeness (fraction of recovered input stars) as well as accuracy. As with the other galaxies in the \cchp series, we find high completeness in F814W. Specifically, we find $\sim 90\%$ completeness at 0.5~mag below (fainter than) the TRGB and $\sim 80\%$ completeness 1~mag below the TRGB. We also find that there is no significant average deviation from the input magnitude near the TRGB. In other words, although the scatter of measured magnitudes increases with magnitude due to deterioration in signal-to-noise, there is no significant systematic offset at the magnitude range in question.

Beyond the median difference between input and output magnitudes from artificial stars, it is still possible to have blends, which, being shifted systematically brighter, could obscure the TRGB. We evaluated the possibility of blends following the surface brightness arguments in \cite{1998AJ....115.2459R}. In \citetalias{hatt2018a}, it was found that only the $\sim10\%$ of the NGC\,4424 footprint closest to the galaxy  had a likelihood of $>5\%$ of containing blends of RGB and AGB populations. In both \ngcfourfour and \ngconesix, the surface brightnesses are lower (lower surface density of sources) resulting in the probability of blends being negligible ($<<1\%$).

\subsection{Optimizing the TRGB edge detection}\label{ssec:opt_trgb}

We now investigate how a randomly generated subset of our artificial star luminosity functions---comparable in count to the observed luminosity functions---affects the measurement of the TRGB as a function of the smoothing in the binned luminosity function. For this and previous works in the \cchp series, our smoothing function is GLOESS (Gaussian-windowed, Locally-Weighted Scatterplot Smoothing), which is a non-parametric interpolation technique. For a single iteration, the luminosity function subset of stars is binned, smoothed, and measured with our edge-detector identically to that of the real dataset. For each simulated TRGB detection, we record its location and the level of smoothing, which we label $\sigma_s$. Ten-thousand TRGB simulations were carried out for each smoothing scale in the range of 0.01-0.18~mag in 0.01~mag intervals.

\begin{figure*}
\centering
\advance\leftskip-1.2cm
\includegraphics[angle=0,width=1.15\textwidth]{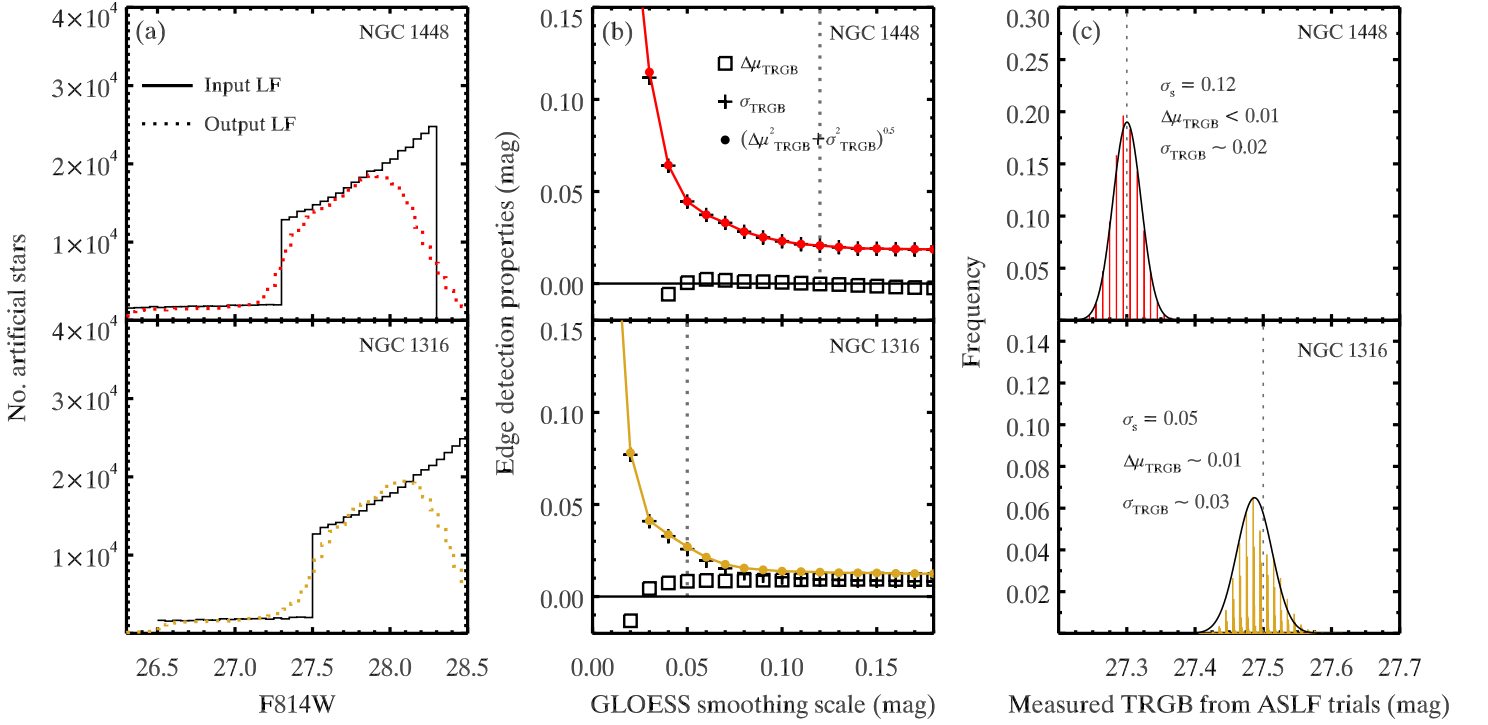}
\caption{Estimating edge detection uncertainties through artificial star tests. Left Panel (a) input (solid) and output (dashed) AGB+RGB artificial star luminosity functions. The input TRGBs for \ngcfourfour and \ngconesix 27.3~mag and 27.5~mag, respectively.
Middle Panel (b) Statistical/random (plus signs), systematic (squares), and combined measurement uncertainties (points and lines) associated with the $[-1,0,+1]$ Sobel edge-detection kernel as a function of GLOESS smoothing scale for each galaxy. Dotted vertical lines mark the chosen level of smoothing for each galaxy. Right Panel (c) The distribution of measured TRGB values at the optimal/chosen level of GLOESS smoothing. 
\label{fig:artstars}}
\end{figure*}

The results of these simulations are given in panels (b) and (c) of \autoref{fig:artstars}. Panel (b) displays the dependency of the average measured location of the TRGB, $\Delta \mu_{\mathrm{TRGB}}$, and the dispersion of measurements, $\sigma_{\mathrm{TRGB}}$, on the level of smoothing of the luminosity function.  Panel (c) displays the dispersion of measured TRGB values from the simulations for the chosen ``optimal'' level of smoothing.

\ngcfourfour and \ngconesix display well-behaved dependencies on smoothing: under-smoothing the luminosity function results in triggering off noise but large levels of smoothing appear to have minimal impact on displacing the location of the input TRGB (low systematic error). For \ngcfourfour, we have chosen a value of $\sigma_s=0.12$~mag to minimize the statistical uncertainty associated with the measurement while being cautious not to over-smooth the data, which in cases of less-idealized datasets could cause competing peaks in the edge-detector to merge, resulting in a robust but displaced (systematically offset) TRGB measurement. The predicted systematic uncertainty for the chosen level of smoothing is a negligible $<0.01$~mag, and at this level, the width of the distribution of measured TRGBs (the statistical uncertainty) is $\sim\mTRGBngcfourfourSTATerrROUNDED$~mag (see \autoref{fig:artstars}c).

In the case of \ngconesix, we have chosen a smaller smoothing scale of $\sigma_s=0.05$~mag because of competing edge detections in the real dataset $\sim0.2-0.3$~mag fainter within the RGB luminosity function. We revisit this observation later in Sections \ref{ssec:dist} and \ref{ssec:ngcdistcompare} and discuss its possible origin. The smoothing scale chosen here is therefore intended to be small enough to resolve the peaks in the edge-detector while being large enough to suppress the likelihood of a false positive in the TRGB detection. 
The anticipated systematic effect on the TRGB measurement is $\sim0.01$~mag while the statistical uncertainty is estimated to be $\sim\mTRGBngconesixSTATerrROUNDED$~mag.

Since AGB stars only blur the TRGB edge-detection, fewer AGB stars than the RGB:AGB ratio assumed here will only decrease the expected measurement uncertainty. To test for an extreme abundance of AGB stars, we decreased the ratio of RGB:AGB to 1:1 (a population at the TRGB that is 50\% AGB), then re-ran our TRGB edge-detection simulations for NGC 1316 as a test case. We found that the systematic uncertainty increased to only $\sim0.02$~mag from virtually no systematic effect. This result suggests that the S/N of the TRGB discontinuity is sufficiently large to retain high measurement precision even in a `worst-case' scenario of RGB:AGB population abundances.

\subsection{TRGB Measurements and Distances}\label{ssec:dist}

\begin{deluxetable*}{ccccccccccc} 
\tabletypesize{\normalsize} 
\tablewidth{0pt} 
\tablecaption{Summary of TRGB distances to \ngcfourfour and \ngconesix\label{tbl:distances}} 
\tablehead{ 
\colhead{Galaxy} &
\colhead{$m_{\mathrm{TRGB}}$\tablenotemark{a}} &
\colhead{$\sigma_{m}$\tablenotemark{b}} &
\colhead{$A_{\mathrm{F814W}}$} &
\colhead{$\left(m-M\right)_0$\tablenotemark{c}} &
\colhead{$\sigma_{stat}$} &
\colhead{$\sigma_{sys}$} &
\colhead{$D$~(Mpc)} &
\colhead{$\sigma_{stat}$} &
\colhead{$\sigma_{sys}$} &
}
\startdata 
\ngcfourfour & \mTRGBngcfourfourROUNDED & \ngcfourfourmTRGBerrCMBROUNDED  & \IextinctionngcfourfourROUNDED & \ngcfourfourDMROUNDED & \ngcfourfourCMBstaterrROUNDED & \ngcfourfourCMBsyserrROUNDED & \ngcfourfourDistROUNDED & \ngcfourfourDiststaterrROUNDED & \ngcfourfourDistsyserrROUNDED\\
\ngconesix & \mTRGBngconesixROUNDED & \ngconesixmTRGBerrCMBROUNDED  & \IextinctionngconesixROUNDED & \ngconesixDMROUNDED & \ngconesixCMBstaterrROUNDED & \ngconesixCMBsyserrROUNDED & \ngconesixDistROUNDED & \ngconesixDiststaterrROUNDED & \ngconesixDistsyserrROUNDED\\
\enddata 
\tablecomments{Uncertainties in the above distance moduli and physical distances are dominated by the TRGB zero-point and are comparable at the $\pm$0.01~mag level of precision.} 
\tablenotetext{a}{F814W}
\tablenotetext{b}{Combined statistical and systematic uncertainties from the edge detection method and calibration to the \hst flight magnitude system.}
\tablenotetext{c}{$M_I^{\mathrm{TRGB}}=\trgblum\pm0.03\pm0.05$~mag}
\end{deluxetable*} 

\begin{figure*}
\centering
\advance\leftskip-1.2cm
\includegraphics[angle=0,width=0.9\textwidth]{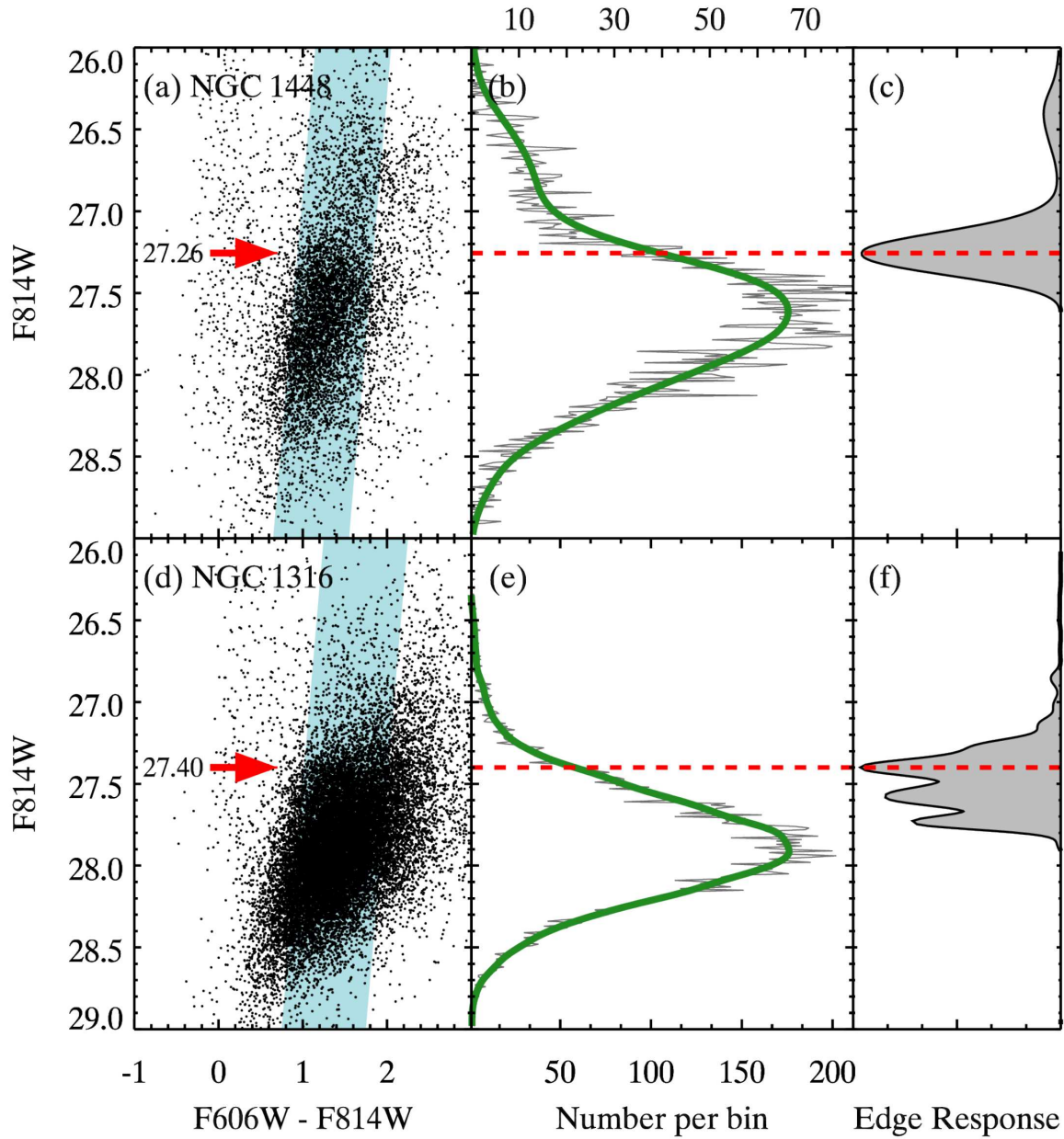}
\vspace{0.0cm}
\caption{TRGB edge detection for \ngcfourfour and \ngconesix. Panels (a) and (d) display the CMDs. A red arrow in each plot shows the location of the measured TRGB. Panels (b) and (e) show the binned luminosity functions in 0.01 mag intervals in gray and the GLOESS-smoothed luminosity functions in solid green. Panels (c) and (f) show the response function of the $[-1,0,+1]$ kernel on the smoothed luminosity functions. These functions are scaled so that their peaks align. A dashed line passes through the location of the greatest change in the luminosity functions. 
\label{fig:distances}}
\end{figure*}

In order to isolate the RGB and achieve the most accurate measurement of the TRGB, we have tested the effect of a color-magnitude selection filter to remove any possible contamination by blue sources that are distinct from the AGB/RGB sequences. 
The slope of the selection windows, $-6$~mag/color, are measured from \cchp ACS/WFC observations of the high S/N and high completeness RGB of M\,31. The left edge of the windows are set to visually rest against the blue edges of the RGB, and we set the red edge of the window to visually encompass the RGB up to $\mathrm{F606W}-\mathrm{F814W}=2.0$, which is  the likely maximum extent of metal-poor stars based on the color uncertainties.
At the chosen levels of GLOESS smoothing, we find that the the value of the TRGB does not differ by more than 0.01~mag from that with no color-magnitude filter. Nevertheless, we apply the color-magnitude selection window for our measurements in the following to ensure any possible systematics are minimized.

The targets of this study were chosen, in part, because of their estimated low foreground extinction (i.e., $E(B-V)\leq0.02$ \citep[][obtained via NED]{2011ApJ...737..103S}. Adopting a \cite{1989ApJ...345..245C} reddening law, the predicted foreground reddening values are $A_{\mathrm{F814W}}=\IextinctionngcfourfourROUNDED$~mag and $A_{\mathrm{F814W}}=\IextinctionngconesixROUNDED$~mag for \ngcfourfour and \ngconesix, respectively. The uncertainty in $E(B-V)$ is estimated to be $\pm0.03$~mag \citep{1998ApJ...500..525S}, which suggests that the foreground extinction for each of these galaxies are statistically consistent with zero. We conservatively include half of the value of the estimated reddening as an additional systematic error in the distance moduli derived below.

Although the foreground reddening is predicted to be very low, it is not yet possible to assess whether there is extinction intrinsic to the halos themselves. One possible test for the presence of halo dust, however, is to  observe whether or not the apparent TRGB magnitude changes with increasing distance into the halo. We tested this possibility by dividing the images into two distinct regions having equal numbers of stars. We re-ran our TRGB simulations with the adjusted star counts for this new test to find the appropriate level of GLOESS smoothing to minimize the combined measurement uncertainties. We found that even with the reduced statistics, the required level of smoothing is comparable to its original value that was found for the full catalogs. In the case of \ngcfourfour, the measured TRGB for the region closest to the galaxy agreed with the further region to within 0.01~mag. For \ngcfourfour, we thus conclude that there is insufficient evidence for halo dust within the current observation footprint.

In the case of possible halo reddening for \ngconesix, the difference in measured TRGB widens to $\sim0.2$~mag, where secondary peaks from the edge-detector (described in Section \ref{ssec:opt_trgb}) become the dominant signals. This difference could suggest the existence of halo reddening, but by further dividing the imaging into quarters, for example, it is apparent that the measured TRGB values are not a function of radial distance from the galaxy. Instead, the stars contributing to the peaks are spread approximately equally across the entire footprint, and the shift in dominance between the edge detection peaks appears to be due to fluctuations in population counts. In Section \ref{ssec:ngcdistcompare}, we discuss the possible origin of the additional peaks within the RGB. Since the additional peaks appear to lie firmly within the RGB itself, we associate only the first (and most prominent peak with the full dataset) with the TRGB. In the measurement of \ngconesix TRGB below, we note that we exclude a small region occupied by a recently discovered dwarf galaxy, Fornax UFD1, belonging to \ngconesix \citep{2017ApJ...835L..27L}, which covers only $0.07\%$ of the footprint.

We turn now to the question of metallicity. At high metallicity, a downward sloping TRGB is observed in color-magnitude space for the reddest stars at optical wavelengths, although this effect is greatly diminished in the $I$-band/F814W. In addition, the observations used in this study were specifically crafted to target the metal-poor halos of these galaxies. As a result, given that the TRGBs in our sample of galaxy halos do not show any discernible color-magnitude dependence, we do not apply color-magnitude ``rectification'' tools \citep[for ACS filters see][]{2017ApJ...835...28J}.

\autoref{fig:distances} displays the results of the TRGB measurement using the optimally selected GLOESS-smoothing scales for each galaxy. 
We find the following F814W ($I$-band equivalent) TRGB magnitudes: for \ngcfourfour, $I(TRGB) = $ $\mTRGBngcfourfourROUNDED\pm\ngcfourfourmTRGBerrCMBROUNDED$~mag, and for \ngconesix, $I(TRGB) = $ $\mTRGBngconesixROUNDED\pm\ngconesixmTRGBerrCMBROUNDED$~mag. These uncertainties combine both the statistical and systematic associated with the calibration and measurement. As with previous papers in this series, we adopt a provisional zero-point for the $I$-band/F814W TRGB based on the geometric distance to the Large Magellanic Cloud and an average of published $I$-band TRGB magnitudes in the literature. Adopting $E(B-V) =  0.03\pm0.03$~mag for the reddening of the LMC TRGB stars \citep[see][]{2018arXiv180301277H}, we derive $M_I =-4.00$~mag $\pm 0.03_{stat}\pm 0.05_{sys}$. A summary of the TRGB measurements and extinction-corrected distances is given in \autoref{tbl:distances}. In the following section, we compare our findings with previously determined estimates.

In the previous section, we derived measurement uncertainty estimates for anticipated TRGB values of 27.3 and 27.5~mag for \ngcfourfour and \ngconesix, respectively, which  we note are not identical to those measured with the real datasets. The major driving factor in how well the TRGB is measured for a given amount of data smoothing is the typical photometric uncertainties of stars at the TRGB interface. Since the spread in photometric uncertainties for stars near the anticipated TRGBs is greater than the difference between our simulated and measured TRGB values, we expect no significant change in the estimated TRGB measurement uncertainty should we have generated new artificial stars and simulations.

\section{Distance comparisons}\label{sec:dist_compare}

\begin{figure*}
\centering
\includegraphics[angle=0,width=1.0\textwidth]{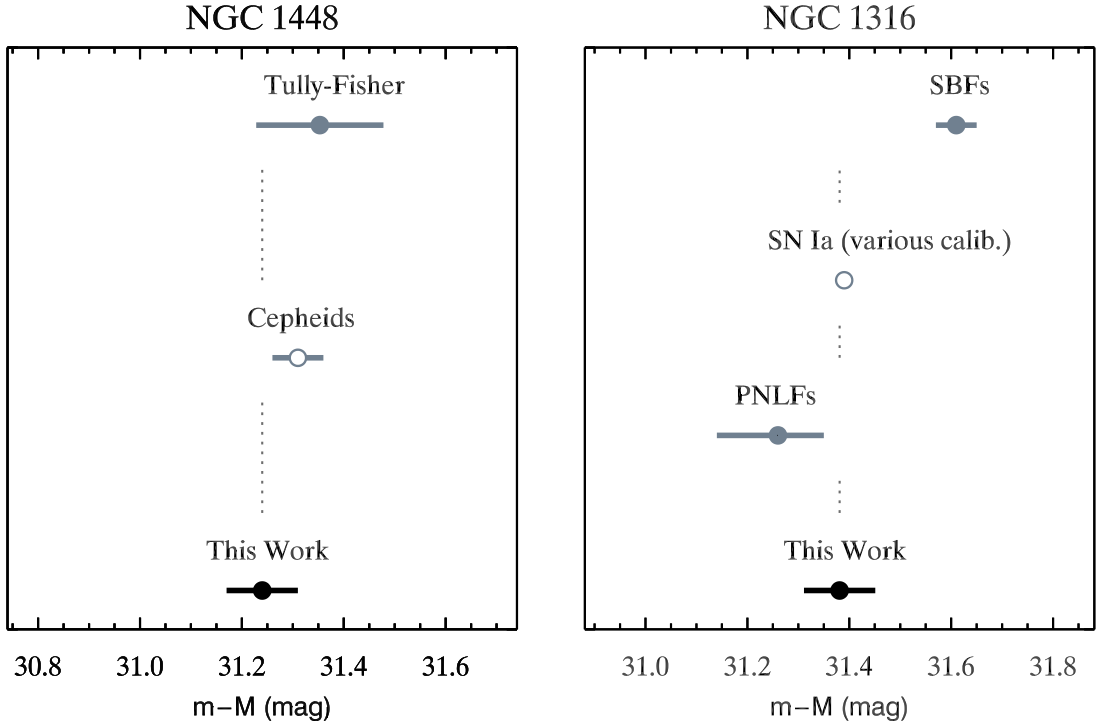}
\vspace{0.0cm}
\caption{Distance moduli and their uncertainties as part of this work compared with the existing literature (statistical and systematic errors are added in quadrature). The Cepheid distance (only available for \ngcfourfour) is published as ``approximate'' and is denoted here by an open circle \citep[see Table 5 of][]{2016ApJ...826...56R}. SBFs, PNLFs, and the \sne of \ngconesix are the only direct and reliable distance determinations to the galaxy (others report distances to the Fornax Cluster as a proxy). The \sne average is not an optimal comparative tool for our measurement given the calibration goals of the \cchp, but their average is presented here as a reference point (also denoted by an open circle) to compare against the large difference with SBFs. All measurements from the literature are taken at face value (non-adjusted for zero-points, extinction, etc.) and the displayed uncertainties are their respective errors on the mean. Vertical dotted lines pass through the results of this work. 
\label{fig:distance_comparisons}}
\end{figure*}

\subsection{\ngcfourfour}

Distance estimates for \ngcfourfour are based primarily on Tully-Fisher, \sne or SNe~II. There are 2 `modern' (since the year 2000) publications at the time of writing for the Tully-Fisher method \citep[][]{2009ApJS..182..474S,2016AJ....152...50T}, and we have used the publications that make use of the most current observations that are part of surveys. The Tully-Fisher relation produces an average distance modulus of $31.35\pm0.12$~mag. We have taken these published distances at face value and have not adjusted in any way for their assumptions on zero-point, extinction, etc. The Tully-Fisher method is in good agreement with our distance modulus of $\ngcfourfourDMROUNDED\pm\ngcfourfourCMBstaterrROUNDED_{stat}\pm\ngcfourfourCMBsyserrROUNDED_{sys}$~mag (1 standard deviation of the average error in distance).

Regarding the supernovae distances, a direct comparison is not entirely suitable here given the \sne calibration objectives of the \cchp. Nonetheless, we report their average distance since they are a considerable component of the literature for distances to \ngcfourfour. Among several publications, there is 1 publication with unique observations of the \sne \citep[][]{2003AJ....125..166K}, which gives an $H$-band distance modulus of $31.04\pm0.14$~mag and a $BVI$ distance of $31.29\pm0.08$~mag. The $H$-band distance appears to be only in approximate agreement ($\sim$2-$\sigma$ of their reported uncertainties), but the optical distance in good agreement with our TRGB distance to within the mutual reported errors. 

Finally, a single Cepheid-based distance exists for \ngcfourfour. \cite{rie16} quote an ``approximate'' Cepheid distance of $31.31\pm0.05$~mag. This distance is in agreement with ours to within 1.4-$\sigma$ of their reported uncertainties, and our measurement is in agreement with theirs to within 1-$\sigma$. 

\subsection{\ngconesix}\label{ssec:ngcdistcompare}

The variety of distance determinations that exist for \ngconesix include Tully-Fisher, Faber-Jackson, Surface Brightness Fluctuations (SBFs), Planetary Nebula Luminosity functions (PNLFs), Globular Cluster Luminosity Functions (GCLFs), and Cepheids, among others. The numerous methods that derive distances for \ngconesix are largely or exclusively derived for the Fornax Cluster, however, not \ngconesix itself, and are therefore not ideal for a comparison here against our direct measurement of the galaxy.

We have endeavored to select only those distances that pertain exclusively to \ngconesix. Regarding GCLFs, attempts have either resulted in poor fits \citep{2010ApJ...717..603V} or have been excluded from analysis because of peculiarities in their size distributions \citep{2010ApJ...715.1419M}. Modern estimates from PNLFs, on the other hand, yield a distance modulus of $31.26\substack{+0.9 \\ -0.12}$~mag \citep{2007ApJ...657...76F}, which based on their estimated errors is only approximately 1-$\sigma$ from our measurement. Next, SBFs produce an average distance modulus of $31.61\pm0.04$~mag \citep{2001ApJ...546..681T,2001ApJ...559..584A,2003ApJ...583..712J,2007ApJ...668..130C,2009ApJ...694..556B,2010ApJ...724..657B,2013A&A...552A.106C}, where the median distance and median uncertainty for multiple measurements in a single publication are taken as representative. 

The average distance from SBFs is $\sim0.2$~mag fainter than our reported distance modulus, which is a substantial difference, but this average distance \emph{is} curiously more closely associated with one of the distances that one of the secondary peaks in the edge-detector would yield, $\sim31.6-31.7$~mag, under the assumption that it represents a TRGB measurement. This alignment is perhaps not a coincidence and suggests that different underlying populations along the line-of-sight of \ngconesix could be affecting accurate measurements with SBFs where the populations cannot be distinguished, unlike this work where the populations can be distinguished using a CMD. To add weight to the argument of a distinct underlying population, either within \ngconesix or along its line-of-sight, the greater distance from SBFs closely aligns with the average distance to the Fornax Cluster via, for example, Cepheids \citep[$31.60\pm0.04$~mag,][]{2000ApJ...529..745F}. Although \ngconesix is at the relative edge of the Cluster, it is bordered by several members that could have played a role in its merger history and may possibly contribute to the stellar populations along the line-of-sight.
Furthermore, evidence of additional stellar populations within \ngconesix, possibly from merger events, have been suggested in recent articles \citep[e.g. the presence of young globular clusters and field stars by][]{2018MNRAS.tmp.1362S}.

The third direct distance measurement to \ngconesix (i.e., not relying on the Fornax Cluster as a whole) is through its \sne. Though not an ideal comparison because of the \sne calibration goals of the \cchp, we include a discussion here for a broader reference point to our measurement. The SNe Ia for \ngconesix produce an average distance modulus of $31.39\pm0.01$~mag \citep[the most comprehensive review having been done by][]{2010AJ....140.2036S}, and our distance of $\ngconesixDMROUNDED\pm\ngconesixCMBstaterrROUNDED_{stat}\pm\ngconesixCMBsyserrROUNDED_{sys}$~mag agrees to within a single standard deviation. The good agreement with the \sne distances, whose calibration is set by other methods and galaxies (often Cepheids), is reassuring in light of the large difference with SBFs.

Finally, the fourth distance to \ngconesix comes from the recently discovered dwarf galaxy Fornax UFD1, also using the observations that are analyzed in this work. Its TRGB distance was measured to be $31.35\pm0.15$~mag \citep{2017ApJ...835L..27L}, where we have adjusted their value for $M_I$ \citep[quoted in][]{2017ApJ...835...28J} to ours. The relatively small number of stars contributing to its TRGB inflates the uncertainty in its distance, but there is mutual agreement between the distances to within 1-$\sigma$ of the quoted errors.

\section{Conclusions}\label{sec:conc}

We have determined the first Tip of the Red Giant Branch distances for two \sne-host galaxies, \ngcfourfour and \ngconesix, which are an integral part of an on-going effort by the \cchp to independently establish the \sne distance scale using Population~II stars.

We find good agreement between these latest (and systematically independent) results in comparison to a variety of previously published distances for each of these galaxies. Moreover, the TRGB distances determined here are higher-precision than most existing estimates. Of publications that report smaller uncertainties in distance, most are derived from the \sne themselves that are inherently linked to a zero-point with an indeterminate or unreported systematic uncertainty. 
For this reason, the results presented here will serve as valuable, independent calibrators for the \sne extragalactic distance scale for the foreseeable future. 

With future \emph{Gaia} data releases, the \cchp will be refining the TRGB distance scale locally using Milky Way RGB stars, thereby improving the zero-point accuracy of the TRGB method as a whole and further improving upon the systematics of the distances estimates reported here. Longer-term, with the pending launch of \emph{JWST}, the galaxies studied here will serve as nearby rungs on the ever-increasing distance scale being built upon the TRGB method.

\section*{Acknowledgments}
We thank Peter Stetson for a copy of $\textsc{DAOPHOT}$ as well as his helpful engagement on its usage. 
Support for this work was provided by NASA through Hubble Fellowship grant \#51386.01 awarded to R.L.B. by the Space Telescope Science Institute, which is operated by the Association of  Universities for Research in Astronomy, Inc., for NASA, under contract NAS 5-26555.
Authors MGL and ISJ were supported by the National Research Foundation of Korea (NRF) grant funded by the Korea Government (MSIP) No. 2017R1A2B4004632.
Support for program \#13691 was provided by NASA through a grant from the Space Telescope Science Institute, which is operated by the Association of Universities for Research in Astronomy, Inc., under NASA contract NAS 5-26555.
This research has made use of the NASA/IPAC Extragalactic Database (NED), which is operated by the Jet Propulsion Laboratory, California Institute of Technology, under contract with the National Aeronautics and Space Administration.

\facility{HST (ACS/WFC)}
\software{DAOPHOT \citep{1987PASP...99..191S}, ALLFRAME \citep{1994PASP..106..250S}, TinyTim \citep{2011SPIE.8127E..0JK}}
\vfill\eject


\end{document}